\documentclass[aps,prd,preprint,nofootinbib]{revtex4}

\usepackage{graphicx}
\usepackage{amsmath}

\def\ltwid{\mathrel{\raise.3ex\hbox{$<$\kern-.75em\lower1ex\hbox{$\sim$}}}}
\def\gtwid{\mathrel{\raise.3ex\hbox{$>$\kern-.75em\lower1ex\hbox{$\sim$}}}}

\def\square{\kern1pt\vbox{\hrule height 1.2pt\hbox{\vrule width 1.2pt\hskip 3pt
   \vbox{\vskip 6pt}\hskip 3pt\vrule width 0.6pt}\hrule height 0.6pt}\kern1pt}
\def\overleftrightarrow#1{\vbox{\ialign{##\crcr
     $\leftrightarrow$\crcr\noalign{\kern-1pt\nointerlineskip}
     $\hfil\displaystyle{#1}\hfil$\crcr}}}
     
\newcommand{\be}{\begin{equation}}
\newcommand{\ee}{\end{equation}}
\newcommand{\bea}{\begin{eqnarray}}
\newcommand{\eea}{\end{eqnarray}}
\newcommand{\nn}{\nonumber}

\begin{document}

\begin{titlepage}

\begin{flushright} 
UFIFT-QG-15-08
\end{flushright}

\vskip 1cm

\begin{center}
{\bf Quantum Scalar Corrections to the Gravitational Potentials on de Sitter Background}
\end{center}

\vskip 1cm

\begin{center}
Sohyun Park$^{1*}$, Tomislav Prokopec$^{2\S}$ and R. P. Woodard$^{3\dagger}$
\end{center}

\begin{center}
\it{$^{1}$ Korea Astronomy and Space Science Institute\\
Daejeon, 305-348, REPUBLIC OF KOREA}
\end{center}

\begin{center}
\it{$^{2}$ Institute of Theoretical Physics, Spinoza Institute and the \\
Center for Extreme Matter and Emergent Phenomena (EMME$\Phi$)\\
Utrecht University, Leuvenlaan 4, 3584 CE Utrecht, THE NETHERLANDS}
\end{center}

\begin{center}
\it{$^{3}$ Department of Physics, University of Florida \\
Gainesville, FL 32611, UNITED STATES}
\end{center}

\vskip .5cm

\begin{center}
ABSTRACT
\end{center}

We employ the graviton self-energy induced by a massless, minimally coupled (MMC)
scalar on de Sitter background to compute the quantum corrections to 
the gravitational potentials of a static point particle with a mass $M$. 
The Schwinger-Keldysh formalism is used to derive real and causal effective 
field equations. When evaluated at the one-loop order, the gravitational 
potentials exhibit a secular decrease in the observed gravitational coupling $G$.
This can also be interpreted as a (time dependent) anti-screening of the mass $M$.

\begin{flushleft}
\end{flushleft}

\vskip 1cm

\begin{flushleft}
$^*$ e-mail: spark1@kasi.re.kr \\
$^{\S}$ e-mail: T.Prokopec@uu.nl\\
$^{\dagger}$  e-mail: woodard@phys.ufl.edu
\end{flushleft}

\end{titlepage}

\section{Introduction}\label{intro}

Studies of quantum loop corrections to the gravitational potentials in flat space
background have a long history~\cite{AFR,CDH,CD,DMC1,MJD,HL,DL,JFD,MK,ABS,KK,gravpot,DM,SMA}.
These studies are typically based on computing the scattering amplitude for two
massive particles and then solving the inverse scattering problem to reconstruct 
a Newtonian potential which would produce the same scattering amplitude in quantum
mechanics. This technique is well-tested and has the tremendous advantage of being
independent of the choice of gauge and of field variable. However, it seems artificially
restricted to asymptotic scattering problems, as compared with the time-dependent 
effects which can be explored using the classical field equations. And its application
to cosmology seems inappropriate because the formal S-matrix which can be defined for 
massive scalars on de Sitter \cite{Marolf:2012kh} is not observable.
  
A more suitable technique for time-dependent sources and cosmological backgrounds is 
the Schwinger-Keldysh, or in-in, formalism~\cite{earlySK}, which provides expectation 
values of operators instead of in-out matrix elements. The authors have previously 
solved the Schwinger-Keldysh effective field equations to work out quantum corrections
(from a massless, minimally coupled (MMC) scalar) to the two potentials of a point mass 
on flat space background~\cite{PWforceflat,MP}. When graviton and gauge particles start 
to appear in loops the problem of gauge dependence must be faced, but that is not an 
issue here. And it should be noted that the Schwinger-Keldysh results are consistent 
with those derived using conventional scattering techniques. They also furnish an 
essential correspondence limit for the current computation in de Sitter. 

De Sitter space is of particular interest in cosmology as a paradigm for the
background of primordial inflation. A generic prediction of inflation is that the
quantum fluctuations of MMC scalars and gravitons are amplified and preserved to late 
times so that they seed large scale structure formation \cite{Starobinsky:1979ty,
Mukhanov:1981xt,Hawking:1982cz,Guth:1982ec,Starobinsky:1982ee,Bardeen:1983qw,
Mukhanov:1985rz,Mukhanov:1990me}. This is a tree order effect, but the same quantum 
fluctuations inevitably give rise to loop effects which have been studied in recent 
years \cite{phi4,SQED,PP,Yukawa,mw,KW1,kahya,PW,LW,PMTW,LPPW,BKP,CV,FRV,FPRV,DS,UM}. 
The purpose of this paper is to learn how a loop of MMC scalars changes the gravitational 
potentials of a point mass on de Sitter. This involves three tasks:
\begin{enumerate}
\item{Compute and renormalize the one-loop contribution to the
graviton self-energy $-i[^{\mu\nu}\Sigma^{\rho\sigma}](x;x')$ from
a MMC scalar on de Sitter background;}
\item{Convert the in-out self-energy to the retarded one of the 
Schwinger-Keldysh formalism,}
\begin{equation}
\bigl[{}^{\mu\nu}\Sigma^{\rho\sigma}\bigr](x;x') \rightarrow \bigl[{}^{\mu\nu}\Sigma_{\text{Ret}}^{\rho\sigma}\bigr](x;x') 
\,;
\end{equation}
\item{Solve the quantum corrected, linearized Einstein field equation 
}
\begin{equation}
\mathcal{D}^{\mu\nu\rho\sigma} \kappa h_{\rho\sigma}(x) - \int d^4x' \bigl[{}^{\mu\nu}\Sigma_{\text{Ret}}^{\rho\sigma}\bigr](x;x') \kappa
h_{\rho\sigma}(x') = 8\pi G \mathcal{T}_{\text{lin}}^{\mu\nu}(x)
\,. \label{lineffeq}
\end{equation}
Here $\mathcal{D}^{\mu\nu\rho\sigma} \kappa h_{\rho\sigma}(x)$ is derived by 
expanding the gravitational side of the Einstein equation, $(R^{\mu\nu} + \Lambda 
g^{\mu\nu} -\frac{1}{2}g^{\mu\nu}R) \sqrt{-g}$, about de Sitter background to first
order in the metric perturbation, $\delta g_{\rho\sigma}(x) \equiv a^2(t) \kappa 
h_{\rho\sigma}(x)$, where $a = a(t)$ is the scale factor, $\kappa^2 = 16\pi G$ is 
the quantum gravitational loop counting parameter, $G$ is the Newton constant, 
and $\mathcal{T}_{\text{lin}}^{\mu\nu}(x)$ is the linearized stress-energy 
tensor density. 
\end{enumerate}

The first two steps have been already performed in the Ref. \cite{LPPW} and we 
summarize the results in section II. Section III is devoted to the last step, 
that is to solving the Schwinger-Keldysh effective field equations for the graviton 
field sourced by a static point mass. Our discussion comprises section IV, and some
tedious technical details from section III have been subsumed to an appendix.

\section{Schwinger-Keldysh Effective Field Equations}

The point of this section is to present the Schwinger-Keldysh effective field 
equations which we will solve in the next section. We first set up the background 
geometry and define the graviton field as a perturbation around this background. 
We then give the in-out effective field equations derived in \cite{LPPW} and 
discuss how to solve them perturbatively. Finally, we explain why it is more
appropriate to convert to in-in equations for cosmological backgrounds such as
de Sitter, and we make the conversion.

\subsection{Preliminaries}

We consider the Lagrangian of gravity plus a MMC scalar, 
\bea
\label{Lagrangian}
\mathcal{L} = 
 \frac{1}{16 \pi G} (R-2\Lambda )\sqrt{-g}
- \frac12 \partial_{\mu} \phi \partial_{\nu} \phi g^{\mu\nu} \sqrt{-g}  \;,
\eea
where $G$ is Newton's constant, $R$ is the Ricci scalar and $\Lambda$ is the 
cosmological constant. Our computation is based on perturbation theory in
the Poincar\' e patch of de Sitter space
\be
ds^2 = \hat{g}_{\mu\nu}dx^{\mu}dx^{\nu} 
= a^2(\eta) \eta_{\mu\nu} dx^{\mu}dx^{\nu} \;. 
\label{bkgd_geometry}
\ee
The coordinate ranges are
\be
-\infty < x^0 \equiv \eta < 0 \quad , \quad -\infty < x^i <
+\infty \; .
\ee
Here the scale factor depends on conformal time $\eta$ as, $a = -1/H\eta$ and 
the Hubble parameter $H = \sqrt{\frac13 \Lambda}$ is constant. It is also useful 
to employ the de Sitter length function, 
\begin{equation}
y(x;x') \equiv H^2 a a' \Bigl[ \Vert \vec{x} \!-\! \vec{x}' \Vert^2
- (\vert \eta \!-\! \eta'\vert - i \varepsilon )^2 \Bigr] \; ,
\label{ydef}
\end{equation}
where $a \equiv a(\eta)$ and $a' \equiv a(\eta')$. Note that $y(x;x')$ is 
related to geodesic distance on de Sitter $\ell(x;x')$ as,
$y(x;x')|_{\epsilon=0}=4\sin^2(H\ell(x;x')/2)$.
We define the graviton field $h_{\mu\nu}$ by subtracting the background from
the full metric and then conformally rescaling,
\be
h_{\mu\nu}(x) \equiv \frac{g_{\mu\nu}(x) \!-\!
\hat{g}_{\mu\nu}(x)}{\kappa a^2}  
\quad \mbox{or} 
\quad g_{\mu\nu}(x) = a^2 \Bigl[\eta_{\mu\nu} + \kappa h_{\mu\nu}(x)\Bigr] 
\equiv a^2{\widetilde g}_{\mu\nu}(x) \;,
\label{htochi}
\ee
where $\kappa^2 \equiv 16\pi G$ is the loop-counting parameter of quantum gravity.

\subsection{Effective field equations}

Varying the one-particle irreducible (1PI) effective action corresponding to the Lagrangian \eqref{Lagrangian} with respect to the graviton field $h_{\mu\nu}$, and retaining only
the linear terms gives,
\be
\mathcal{D}^{\mu\nu\rho\sigma} h_{\rho\sigma}(x) - \int d^4x' \bigl[{}^{\mu\nu}\Sigma^{\rho\sigma}\bigr](x;x') h_{\rho\sigma}(x') 
= \frac{\kappa}{2}\mathcal{T}_{\text{lin}}^{\mu\nu}(x) \;.
\label{lineffeq2}
\ee
Here the Lichnerowicz operator for de Sitter is~\cite{WW,LPPW},
\bea
\lefteqn{
\mathcal{D}^{\mu\nu\rho\sigma} = \frac12 a^2 \Bigl[ \left(\eta^{\mu(\rho}\eta^{\sigma)\nu} - \eta^{\mu\nu}\eta^{\rho\sigma}\right)\partial^2 + \eta^{\mu\nu}\partial^{\rho}\partial^{\sigma} +\eta^{\rho\sigma}\partial^{\mu}\partial^{\nu} - 2\partial^{(\mu}\eta^{\nu)(\rho}\partial^{\sigma)}  \Bigr]
} \nonumber \\
& & \hspace{1cm}
+Ha^3\Bigl[ \left( \eta^{\mu\nu}\eta^{\rho\sigma} - \eta^{\mu(\rho}\eta^{\sigma)\nu} \right)\partial_0 - 2\eta^{\mu\nu}\delta_0^{(\rho}\partial^{\sigma)} + 2\delta_0^{(\rho}\eta^{\sigma)(\mu}\partial^{\nu)} \Bigr] +3H^2a^4 \eta^{\mu\nu}\delta_0^{\rho}\delta_0^{\sigma} \; . 
\label{Lich}
\eea
Quantum corrections come from the graviton self-energy whose general form is,
\bea
\lefteqn{-i\Bigl[ \mbox{}^{\mu\nu} \Sigma^{\rho\sigma}\Bigr](x;x') =
\mathcal{F}^{\mu\nu}(x) \times \mathcal{F}^{\rho\sigma}(x') \Bigl[F_0(x;x') \Bigr] 
+ \mathcal{G}^{\mu\nu}(x) \times
\mathcal{G}^{\rho\sigma}(x') \Bigl[ G_0(x;x') \Bigr] } 
\nn\\
& & \hspace{3cm} +
\mathcal{F}^{\mu\nu\rho\sigma} \Bigl[ F_2(x;x') \Bigr] +
\mathcal{G}^{\mu\nu\rho\sigma} \Bigl[ G_2(x;x') \Bigr] \; . 
\qquad \qquad \qquad \qquad
\label{oursigma}
\eea
The four projection operators 
$\mathcal{F}^{\mu\nu}$, $\mathcal{G}^{\mu\nu}$, $\mathcal{F}^{\mu\nu\rho\sigma}$ and
$\mathcal{G}^{\mu\nu\rho\sigma}$
and one loop results for the corresponding structure functions $F_0, G_0, F_2,$ and 
$G_2$ are given in Ref.~\cite{LPPW}.

It is convenient to re-express the action of the Lichnerowicz operator on the 
graviton by extracting the scale factor 
$a$,~\footnote {Due to an error, only the first term in~(\ref{defEmunu}) is found in Ref.~\cite{LPPW}.}
\be
E^{\mu\nu} \equiv \mathcal{D}^{\mu\nu\rho\sigma} h_{\rho\sigma}(x)  
= \partial_{\alpha} \Bigl[a^2\mathcal{L}^{\mu\nu\rho\sigma\alpha\beta}\partial_{\beta} h_{\rho\sigma}(x) \Bigr] 
+ \partial_{\alpha} \Bigl[Ha^3 \eta^{\mu\nu} h^{\alpha0} \Bigr] - Ha^3\eta^{0(\mu}\partial^{\nu)}h \;,
\label{defEmunu}
\ee 
where the Lichnerowicz tensor factor $\mathcal{L}^{\mu\nu\rho\sigma\alpha\beta}$ is
\be
\mathcal{L}^{\mu\nu\rho\sigma\alpha\beta} \equiv \frac12
\eta^{\alpha\beta} \Bigl[ \eta^{\mu (\rho} \eta^{\sigma) \nu} \!-\!
\eta^{\mu\nu} \eta^{\rho\sigma} \Bigr] + \frac12 \eta^{\mu\nu} \eta^{\rho (\alpha}
\eta^{\beta) \sigma} + \frac12 \eta^{\rho\sigma} \eta^{\mu (\alpha}
\eta^{\beta) \nu} - \eta^{\alpha (\rho} \eta^{\sigma) (\mu}
\eta^{\nu) \beta}\;.
\ee
For the quantum correction, we extract the unprimed derivatives from the $x'^{\mu}$ 
integration and partially integrate the primed derivatives, bringing the effective 
field equation \eqref{lineffeq2} to the form
\begin{eqnarray}
\lefteqn{ E^{\mu\nu}
= \frac{\kappa}{2}\mathcal{T}_{\text{lin}}^{\mu\nu}(x)
+ \mathcal{F}^{\mu\nu} \!\! \int \!\! d^4x' \,iF_0(x;x') \widetilde{R}_{\rm lin}(x') 
+ \mathcal{G}^{\mu\nu} \!\! \int \!\! d^4x' \, iG_0(x;x') \overline{ \widetilde{R}}_{\rm lin}(x') 
} 
\nonumber \\
& & \hspace{1cm} - 2 \partial_{\alpha} \partial_{\beta} \!\! \int
\!\! d^4x' \Bigl[ iF_2(x;x') \widetilde{C}^{ \mu\alpha\nu\beta}_{\rm
lin}(x') + iG_2(x;x') \overline{ \widetilde{C}}^{~
\mu\alpha\nu\beta}_{\rm lin}\!(x') \Bigr] \nonumber \\
& & \hspace{1.5cm} + \Bigl[\eta^{\mu\nu} \partial_k \partial_{\ell}
\!-\! 2 \delta^{(\mu}_{(k} \partial^{\nu )} \partial_{\ell )} \!+\!
\delta^{(\mu}_k \delta^{\nu )}_{\ell} \partial^2 \Bigr] \!\!
\int \!\! d^4x' i G_2(x;x') \widetilde{C}^{0k0\ell}_{\rm lin}(x')
\; . \qquad 
\label{goodeqn}
\end{eqnarray}
Here $\widetilde{R}_{\rm lin}$ and $\widetilde{C}^{\alpha\beta\gamma\delta}_{\rm lin}$
are the linearized Ricci scalar and Weyl tensor of the conformally rescaled metric.
$\overline{ \widetilde{R}}_{\rm lin}$ and $\overline{\widetilde{C}}_{\rm lin}^{~\alpha\beta\gamma\delta}$ are their purely spatial parts, respectively. 

\subsection{Perturbative solution}

Because we only possess one loop results for the structure functions, we must solve 
\eqref{goodeqn} perturbatively by expanding the graviton field and the structure 
functions in loop orders, 
\begin{eqnarray}
h_{\mu\nu}(x) & = & h^{(0)}_{\mu\nu}(x) + h^{(1)}_{\mu\nu}(x) + h^{(2)}_{\mu\nu}(x) + \dots \qquad 
\label{h-exp}
\\
F_{0,2}(x;x') & = & 0 + F^{(1)}_{0,2}(x;x') + F^{(2)}_{0,2}(x;x') + \dots \qquad 
\label{F-exp}
\\
G_{0,2}(x;x') & = & 0 + G^{(1)}_{0,2}(x;x') + G^{(2)}_{0,2}(x;x') + \dots \qquad
\label{G-exp}
\end{eqnarray}
By substituting (\ref{h-exp} - \ref{G-exp}) into \eqref{goodeqn}, we obtain equations
for the tree order field $h^{(0)}_{\mu\nu}$ and the one-loop field $h^{(1)}_{\mu\nu}$,

\bea
\lefteqn{ {E^{\mu\nu}}^{(0)}(x) =\frac{\kappa}{2}\mathcal{T}_{\text{lin}}^{\mu\nu}(x)
} 
\label{0loopeqn}
\\
\lefteqn{{E^{\mu\nu}}^{(1)}(x)
 = \mathcal{F}^{\mu\nu} \!\! \int \!\! d^4x' \,
iF^{(1)}_0(x;x') \widetilde{R}_{\rm lin 0}(x') + \mathcal{G}^{\mu\nu} \!\! \int \!\!
d^4x' \, iG^{(1)}_0(x;x') \overline{ \widetilde{R}}_{\rm lin 0}(x')} 
\nn \\
& & \hspace{1cm} - 2 \partial_{\alpha} \partial_{\beta} \!\! \int
\!\! d^4x' \Bigl[ iF^{(1)}_2(x;x') \widetilde{C}^{ \mu\alpha\nu\beta}_{\rm lin 0}(x') 
+ iG^{(1)}_2(x;x') \overline{ \widetilde{C}}^{~\mu\alpha\nu\beta}_{\rm lin 0}\!(x') \Bigr] 
\nn\\
& & \hspace{1.5cm} + \Bigl[\eta^{\mu\nu} \partial_k \partial_{\ell}
\!-\! 2 \delta^{(\mu}_{(k} \partial^{\nu )} \partial_{\ell )} \!+\!
\delta^{(\mu}_k \delta^{\nu )}_{\ell} \partial^2 \Bigr] \!\!
\int \!\! d^4x' i G^{(1)}_2(x;x') \widetilde{C}^{0k0\ell}_{\rm lin 0}(x')
\equiv \mathcal{S}^{\mu\nu}(x)
\;. \quad \quad 
\label{1loopeqn}
\eea
Here ${E^{\mu\nu}}^{(\ell)} \equiv \mathcal{D}^{\mu\nu\rho\sigma} h^{(\ell)}_{\rho\sigma}$.
Note that in \eqref{0loopeqn} we regard the matter source as 0th order, assuming the 
stress tensor includes no loop corrections from the 1PI 1-point function. The solution 
of the 0th order equation $h^{(0)}_{\mu\nu}$ enters the right hand side of the 1st 
order equation \eqref{1loopeqn} to provide sources for the one-loop field $h^{(1)}_{\mu\nu}$.

\subsection{Schwinger-Keldysh formalism}

The perturbative effective field equation \eqref{1loopeqn} seems to be ready for use, but
if one were to interpret it in the spirit of the {\it in-out} formalism it would possess 
two disturbing features:
\begin{itemize}
\item{{\bf Acausality:}  The in-out effective field equation at $x^{\mu}$ receives influence
from points $x'^{\mu}$ which lie in the future of $x^\mu$, and at spacelike separation from it.}
\item{{\bf Imaginary parts:} The in-out effective field develops an imaginary part if there
is particle production.}
\end{itemize}
Neither of these features prevents one from describing flat space scattering problems, but they would be problematic for cosmological settings in which we do not know what happens in the asymptotic future and the more natural question is how the fields evolve when released at 
finite time in some prepared state. That question is answered by the Schwinger-Keldysh formalism \cite{earlySK}. This technique produces true expectation values, rather than in-out matrix 
elements, so the effective field equations at $x^{\mu}$ depend only on points $x'^{\mu}$ on or within its past light-cone, and the effective fields associated with Hermitian operators are real.
Because excellent reviews on the Schwinger-Keldysh formalism exist~\cite{Chou:1984es,Jordan:1986ug,
Calzetta:1986ey,Ford:2004wc}, and the current authors have described it before in \cite{LPPW}, 
we merely comment that the linearized Schwinger-Keldysh effective field equation is obtained by
replacing the in-out self-energy with its retarded counterpart, 
\begin{equation}
\bigl[{}^{\mu\nu}\Sigma^{\rho\sigma}\bigr](x;x') \rightarrow \bigl[{}^{\mu\nu}\Sigma_{\text{Ret}}^{\rho\sigma}\bigr](x;x') \equiv \bigl[{}^{\mu\nu}\Sigma_{++}^{\rho\sigma}\bigr](x;x')+\bigl[{}^{\mu\nu}\Sigma_{+-}^{\rho\sigma}\bigr](x;x') \; . 
\label{SKSE1}
\end{equation} 

\noindent
In this expression, we obtain $\bigl[{}^{\mu\nu}\Sigma_{++}^{\rho\sigma}\bigr]$ and $\bigl[{}^{\mu\nu}\Sigma_{+-}^{\rho\sigma}\bigr]$ from \eqref{oursigma} by replacing the de Sitter length function $y(x;x')$
by $y_{\scriptscriptstyle ++}(x;x')$ and $y_{\scriptscriptstyle +-}(x;x')$, respectively, where
\begin{eqnarray}
y_{\scriptscriptstyle ++}(x;x') &\equiv& H^2 a a' 
\Bigl[ \Vert \vec{x}\!-\! \vec{x}\,' \Vert^2 - (\vert \eta \!-\! \eta' \vert \!-\! i\varepsilon)^2 \Bigr]  = y(x,x')\; , 
\\
y_{\scriptscriptstyle +-}(x;x') &\equiv& H^2 a a' 
\Bigl[ \Vert \vec{x}\!-\! \vec{x}\,' \Vert^2 - (\eta \!-\! \eta' \!+\! i\varepsilon)^2 \Bigr] \; . 
\end{eqnarray}
This converts the nonzero structure functions in \eqref{1loopeqn} to the retarded ones of the Schwinger-Keldysh 
formalism~\cite{LPPW},
\begin{eqnarray}
\lefteqn{ F^{(1)}_{0}(x;x') 
= \frac{i \kappa^2}{576 \pi^3} \Biggl\{
\frac{\partial^4 \!-\! 4 H^2 a a' \partial^2}{16}
\biggl[ \Bigl[ \ln\Bigl( \frac{- y}{4 a a'} \Bigr)
\!-\! 1\Bigr] \Theta \biggr] - \frac14 H^2 a a' \ln(a a')
\partial^2 \Theta } \nonumber \\
& & \hspace{3cm} + H^4 a^2 {a'}^2 \biggl[3 \!-\! \frac1{4 \!-\! y} 
\!+\! \frac34 (2 \!-\! y) \ln\Bigl(\frac{ -y}{4 \!-\!y} \Bigr) \biggr] 
\Theta \Biggr\} , \qquad \qquad
\label{F10} \\
\lefteqn{ F^{(1)}_{2} (x;x') 
= \frac{i \kappa^2}{64 \pi^3} \Biggl\{ \!
\frac{\partial^4 \!+\! 20 H^2 a a' \partial^2}{240}\!
\biggl[ \Bigl[ \ln\Bigl( \frac{- y}{4 a a'} \Bigr) - 1
\Bigr] \Theta \biggr] \!+\! \frac{H^2 a a' \ln(a a')}{12} \, 
\partial^2 \Theta } \nonumber \\
& & \hspace{5.5cm} + H^4 a^2 {a'}^2 \biggl[ \frac{-\frac13}{4 \!-\! y} 
\!-\! \frac16 \ln\Bigl(\frac{ -y}{4 \!-\! y} \Bigr) \biggr] \Theta 
\Biggr\} , \qquad  \qquad
\label{F12} \\
\lefteqn{ G^{(1)}_{2}(x;x') 
= \frac{i \kappa^2}{64 \pi^3} \Biggl\{H^4 a^2 {a'}^2 
\biggl[\frac{\frac43}{4 \!-\! y} \!+\! \frac13 \ln\Bigl(\frac{ -y}{4 
\!-\! y} \Bigr) \biggr] \Theta \Biggr\} . } 
\label{G12}
\end{eqnarray}
(Note that $G^{(1)}_0 (x;x')$ is zero for the MMC scalar at one loop.) 
Here the symbol $\Theta$ stands for the Heaviside step function which ensures causality,
\begin{equation}
\Theta \equiv \theta\Bigl( \Delta \eta - \Vert \vec{x} \!-\!
\vec{x}' \Vert\Bigr) \qquad , \qquad \Delta \eta \equiv \eta \!-\!
\eta' \; , 
\label{Theta}
\end{equation}
and now the $i\epsilon$-prescription can be dropped in $-y(x;x')$ in Eqs.~(\ref{F10}-\ref{G12}),
\begin{equation}
-y(x;x') \rightarrow -y(x;x')|_{\epsilon=0} = H^2 a a' \Bigl[ \Delta \eta^2 - \Vert \vec{x} \!-\!
\vec{x}' \Vert^2 \Bigr] \; .
\end{equation}
Also note that the structure functions are pure imaginary, which makes the effective field equation 
\eqref{1loopeqn} 
manifestly real. Therefore, the resulting Schwinger-Keldysh effective field equation is causal and real as promised.

\section{Quantum Corrected Gravitational Potentials} 

In this section we solve the effective field equations \eqref{0loopeqn} and \eqref{1loopeqn} which we repeat below,
\bea
{E^{\mu\nu}}^{(0)}(x) &=& \frac{\kappa}{2}\mathcal{T}_{\text{lin}}^{\mu\nu}(x)
\label{0loopeqn-simp}
\\
{E^{\mu\nu}}^{(1)}(x) &=& \mathcal{S}^{\mu\nu}(x)
\label{1loopeqn-simp}
\eea
with the retarded structure functions (\ref{F10}--\ref{G12}) for the graviton field. 
We are interested in quantum loop corrections to the gravitational response of a static point mass $M$. 
The 0th order equation \eqref{0loopeqn-simp} determines the classical response to a point particle. 
The 1st order equation \eqref{1loopeqn-simp} leads to the one-loop correction to the classical gravitational potentials.  

\subsection{Classical solutions}

The linearized stress tensor density $\mathcal{T}_{\text{lin}}^{\mu\nu}(x)$ in \eqref{0loopeqn-simp} for a static point mass $M$ on the de Sitter background takes the form \cite{TW},
\be
\mathcal{T}_{\text{lin}}^{\mu\nu}(x) = -a(\eta)\delta^\mu_0\delta^\nu_0M\delta^3(\vec x) \;.
\ee
The symmetries of this system imply a solution of the form,
\be
h_{00}(x) = f_1(\eta, r)\;, \quad 
h_{0i}(x) = \partial_{i}f_2(\eta, r)\;, \quad
h_{ij}(x) = \delta_{ij}f_3(\eta, r)+ \partial_{i}\partial_{j}f_4(\eta, r) \;,
\ee
where $r \equiv \|\vec x\|$. It is convenient to choose the longitudinal (Newtonian) gauge 
$f_2 = 0$ and $f_4 = 0$.~\footnote{Instead of completely gauge fixing, one could have 
employed the gauge invariant formalism analogous to Refs.~\cite{MP}. The final results, 
expressed through the one-loop corrected Bardeen potentials, can be easily related to the 
results obtained here. For the reasons of simplicity we shall not proceed here along this technically more involved path.} In terms of these variables the $E^{\mu\nu}$ of 
expressions~(\ref{0loopeqn-simp},\ref{1loopeqn-simp}) take the form,
\begin{eqnarray}
 E^{00} &=& a^2\bigg\{-3a^2H^2f_1 +(\nabla^2-3aH\partial_0)f_3\bigg\} \;,
\label{classical eom 00} \\
 E^{0i} &=& a^2\partial_i\bigg\{-aHf_1 -\partial_0f_3\bigg\} \;,
\label{classical eom 0i} \\
 E^{ij} &=& a^2\partial_i\partial_j\bigg\{-\frac12f_1 +\frac12f_3\bigg\}
 \nonumber\\
  &+&
  a^2\delta_{ij}\bigg\{\Big(\frac12\nabla^2+aH\partial_0+3a^2H^2\Big)f_1
                     +\Big(\partial_0^2+2aH\partial_0-\frac12\nabla^2\Big)f_3\bigg\}
\,.\quad
\label{classical eom ij}
\end{eqnarray}
The classical solutions of the 0th order equation \eqref{0loopeqn-simp} are
\begin{eqnarray}
   f_1(x)\rightarrow f_1^{(0)}(x) &=& \frac{2GM}{a(\eta)\|\vec x\|}\equiv - 2\phi^{(0)}(x) \;,
\nonumber\\
   f_3(x)\rightarrow f_3^{(0)}(x) &=& \frac{2GM}{a(\eta)\|\vec x\|} \equiv - 2\psi^{(0)}(x) \;,
   \label{classical solutions}
\end{eqnarray}
where $\phi^{(0)}(x)$ and $\psi^{(0)}(x)$ are the usual potentials in the longitudinal gauge.
Note that these classical solutions~(\ref{classical solutions}) are just conformally rescaled potentials of a point mass in Minkowski space.
At the classical level, both $\mathcal{T}_{\text{lin}}$ and $E^{\mu\nu}_0$ with $f^{(0)}_{1,3}$ given by~(\ref{classical solutions}), obey the (covariant) conservation identities, 
$\partial_\mu \mathcal{T}_{\text{lin}}^{\mu\nu}+aH\delta^\nu_0 \eta_{\alpha\beta} \mathcal{T}_{\text{lin}}^{\alpha\beta}=0$
and
$\partial_\mu E^{\mu\nu}_{0}+aH\delta^\nu_0 \eta_{\alpha\beta} E^{\alpha\beta}_{0}=0$.

\subsection{Computing the one-loop source integrals}

The one-loop source terms on the right hand side of \eqref{1loopeqn-simp} in the 3 + 1 decomposition are 
\bea
\mathcal{S}^{00}
 \!&=&\!\frac{\kappa^2M}{2}\mathcal{F}^{00} \!\! \int \!\!\frac{d\eta'}{a(\eta')} \,[iF^{(1)}_0(x,x')]_{\vec x'= 0}
- \frac{\kappa^2M }{3}\nabla^2  \!\! \int\!\! \frac{d\eta'}{a(\eta')}
\Big[iF^{(1)}_2(x,x')\!+\!\frac12iG^{(1)}_2(x,x')\Big]_{\vec x'= 0} \;,
\label{RHS 00}
\\
\mathcal{S}^{0i}
\!&=&\! \frac{\kappa^2M}{2}\mathcal{F}^{0i}  \!\! \int \!\! \frac{d\eta'}{a(\eta')}[iF^1_0(x,x')]_{\vec x'= 0}
+ \frac{\kappa^2M }{3}\partial_{0}\partial_{i} \!\! \int\!\! \frac{d\eta'}{a(\eta')}
\Big[iF^{(1)}_2(x,x')\!+\!\frac12iG^{(1)}_2(x,x')\Big]_{\vec x'= 0}  \;,
\label{RHS 0i}
\\
\mathcal{S}^{ij}
&=& \frac{\kappa^2M}{2}\mathcal{F}^{ij} \!\!
\int \!\! \frac{d\eta'}{a(\eta')} [iF^{(1)}_0(x,x')]_{\vec x'= 0}
\nonumber\\
&&- 2 \partial_{0}^2\!\! \int\!\! d^4x'
\Bigl[F^{(1)}_2(x;x')+\frac{1}{2}G^{(1)}_2(x;x')\Bigr]
\frac{GM}{a'}\bigg[-\partial_i'\partial_j'+\frac13\delta_{ij}{\nabla'}^2\bigg]\frac{1}{\|\vec x\,'\|}
\nonumber \\
&&-\frac{\kappa^2M}{6}
\Bigl[\delta^{ij}\nabla^2 \!-\!\partial^i\partial^j\Bigr]
\int \!\! \frac{d\eta'}{a(\eta')}
\Bigl[F^{(1)}_2(x;x')+\frac{1}{2}G^{(1)}_2(x;x')\Bigr]_{\vec x'= 0} \;,
\label{RHS ij}
\eea
where
\begin{eqnarray}
\mathcal{F}^{00} &=&a^2\Bigl[\nabla^2-3aH\partial_0 +3a^2H^2\Bigr]a^{-2}=\nabla^2-3aH\partial_0 +9a^2H^2 \;,
\label{F 00}\\
\mathcal{F}^{0i} &=&a^2\partial_i\Bigl[-\partial_0 +aH\Bigr]a^{-2}=\partial_i\Bigl[-\partial_0 +3aH\Bigr] \;,
\label{F 0i}\\
\mathcal{F}^{ij} &=&a^2\Bigl[-(\delta_{ij}\nabla^2-\partial_i\partial_j)+\delta_{ij}(\partial_0^2+aH\partial_0 -3a^2H^2)\Bigr]a^{-2} \;,
\nonumber\\
 &=&-(\delta_{ij}\nabla^2-\partial_i\partial_j)+\delta_{ij}(\partial_0^2-3aH\partial_0-3a^2H^2) \;.
\label{F ij}
\end{eqnarray}
One can check that the left hand side of the effective equation with arbitrary functions $f_{1,3}$ obey a conservation identity,
\begin{equation}
   \partial_\mu E^{\mu\nu} + aH\delta^{\nu}_0 \eta_{\alpha\beta}E^{\alpha\beta}=0 \;,
\label{conservation identity}
\end{equation}
which is a consequence of the contracted linear Bianchi identity. Because of the special (transverse) character of 
${\cal F}^{\mu\nu}$ in Eqs.~(\ref{F 00}--\ref{F ij}),
an analogous conservation identity
holds for the right hand side $\mathcal{S}^{\mu\nu}$ with an arbitrary choice of $F_0^{(1)}, F_2^{(1)}$ and $G_2^{(1)}$.
These represent a nontrivial check of our equations. Moreover, these tell us that
the four equations are not independent. One can solve any two equations; the other two follow from
the conservation identities. (Had one proceeded with the gauge invariant formulation, one would need to 
cleverly combine the four equations into two gauge invariant equations, resulting in the two equations for 
gauge invariant scalar potentials.)

There is one ugly term on the right hand side of the second line of the $(ij)$ equation. All
other terms contain only time integrations, but that term requires a three dimensional spatial integration.
It is hence worth spending some effort and analyse all four equations, to see whether we can get rid of the
spatial integration when evaluating the one-loop corrected $f_{1,3}$. These equations can be easily obtained from
the $(00)$, $(0i)$ and $(ij)$ equations given in~Eqs.~(\ref{classical eom 00}--\ref{classical eom ij}) 
and~(\ref{RHS 00}--\ref{F ij}),~\footnote{One can extract
two equations from the $(ij)$ equation by acting
with the projectors, $\delta_{ij}-\partial_i\partial_j/\nabla^2$ and $(1/3)\delta_{ij}-\partial_i\partial_j/\nabla^2$,
which extract
the terms $\propto \delta_{ij}$ and $\partial_i\partial_j$, respectively.}
\begin{eqnarray}
 -3a^2H^2f_1^{(1)} +(\nabla^2-3aH\partial_0)f_3^{(1)} &=& \frac{\kappa^2M}{2a^2}[\nabla^2-3aH\partial_0 +9a^2H^2] \!\! \int \!\!
\frac{d\eta'}{a(\eta')} \,
[iF^{(1)}_0(x,x')]_{\vec x'= 0}
\nonumber\\
&&\!-\frac{\kappa^2M }{3a^2}\nabla^2  \!\! \int\!\! \frac{d\eta'}{a(\eta')}
\Big[iF^{(1)}_2(x,x')\!+\!\frac12iG^{(1)}_2(x,x')\Big]_{\vec x'= 0}
\;, 
\label{one loop  eom 00} \\
-aH f_1^{(1)} -\partial_0f_3^{(1)}
&=& \frac{\kappa^2M}{2a^2}\Bigl[-\partial_0 +3aH\Bigr] \!\! \int \!\! \frac{d\eta'}{a(\eta')}
[iF^{(1)}_0(x,x')]_{\vec x'= 0}
\nonumber\\
&&
\!+\frac{\kappa^2M }{3a^2}\partial_{0}\! \int\!\! \frac{d\eta'}{a(\eta')}
\Big[iF^{(1)}_2(x,x')\!+\!\frac12iG^{(1)}_2(x,x')\Big]_{\vec x'= 0}
\;,
\label{one loop  eom 0i} \\
-\frac12f_1^{(1)} +\frac12f_3^{(1)}&=&\frac{\kappa^2M}{2a^2}\!\!\int \!\! \frac{d\eta'}{a(\eta')} [iF^{(1)}_0(x,x')]_{\vec x'= 0}
\nonumber\\
&&+\frac{2}{a^2} \partial_{0}^2\!\! \int\!\! d^4x'
\Bigl[iF^{(1)}_2(x;x')+\frac{1}{2}iG^{(1)}_2(x;x')\Bigr]
\frac{GM}{a'}\frac{1}{\|\vec x\,'\|}
\nonumber \\
&&+\frac{\kappa^2M}{6a^2}
\int \!\! \frac{d\eta'}{a(\eta')}
\Bigl[iF^{(1)}_2(x;x')+\frac{1}{2}iG^{(1)}_2(x;x')\Bigr]_{\vec x'= 0}
\;,
\label{one loop eom delta ij}\\
\Big(\frac12\nabla^2+aH\partial_0+3a^2H^2\Big)f_1^{(1)}&+&\Big(\partial_0^2+2aH\partial_0-\frac12\nabla^2\Big)f_3^{(1)}
\nonumber\\
&&\hskip -2cm
 =\frac{\kappa^2M}{2a^2}\Big[-\nabla^2+\partial_0^2-3aH\partial_0-3a^2H^2\Big] \!\!
\int \!\! \frac{d\eta'}{a(\eta')} [iF^{(1)}_0(x,x')]_{\vec x'= 0}
\nonumber\\
&&\hskip -2cm
+\frac{\kappa^2M}{6a^2} (\partial_{0}^2\!-\!\nabla^2)\! \int \!\! \frac{d\eta'}{a(\eta')}
\Bigl[iF^{(1)}_2(x;x')+\frac{1}{2}iG^{(1)}_2(x;x')\Bigr]_{\vec x'= 0}
\;.
\label{one loop eom partial ij}
\end{eqnarray}
The third equation~(\ref{one loop eom delta ij}) tell us that in order to determine the gravitational slip
(defined as the difference of the two potentials) one ought to perform both the integrals over time $\eta'$
and space $\vec x'$. It is convenient to define the two source integrals, the one-loop spin zero $S_0^{(1)}$ and spin two, $S_2^{(1)}$,
as follows,
\begin{eqnarray}
 S_0^{(1)}(x) &\equiv& \int \!\! \frac{d\eta'}{a(\eta')} [iF^{(1)}_0(x,x')]_{\vec x'= 0} \;,
\label{S01}\\
 S_2^{(1)}(x) &\equiv& \int \!\! \frac{d\eta'}{a(\eta')}\Bigl[iF^{(1)}_2(x;x')+\frac{1}{2}iG^{(1)}_2(x;x')\Bigr]_{\vec x'= 0} \;,
\label{S21}
\end{eqnarray}
where
\begin{eqnarray}
 [iF^{(1)}_0(x,x')]_{\vec x'= 0}  &=&\!\! -\frac{\kappa^2}{64\times 9\pi^3}\Bigg\{
               \frac{\partial^4\!-\!4aa'H^2\partial^2}{16}\bigg[\bigg(\ln\bigg(\frac{H^2(\Delta\eta^2\!-\!r^2)}{4}\bigg)\!-\!1\bigg)
   \theta(\Delta\eta-r)
\bigg]
\label{F01}
\\
&&\hskip -2cm
 -\frac14H^2aa'\ln(aa')\partial^2\theta(\Delta\eta\!-\!r)
 + H^4a^2{a'}^2\bigg[3\!-\!\frac{1}{4\!-\!y}\!+\!\frac34(2\!-\!y)\ln\bigg(\frac{-y}{4\!-\!y}\bigg)\bigg]\theta(\Delta\eta\!-\!r)
    \Bigg\},
\qquad\nonumber
\end{eqnarray}
\begin{eqnarray}
\Bigl[iF^{(1)}_2(x;x')\!+\!\frac{1}{2}iG^{(1)}_2(x;x')\Bigr]_{\vec x'= 0}&&
\nonumber\\
&&\hskip -3cm
= -\frac{\kappa^2}{64\times9\pi^3}\Bigg\{
               \!\frac{3\partial^4\!+\!60aa'H^2\partial^2}{80}\!\bigg[\bigg(\ln\bigg(\!\frac{H^2(\Delta\eta^2\!-\!r^2)}{4}\bigg)\!-\!1\bigg)
   \theta(\Delta\eta\!-\!r)
\bigg]
\nonumber\\
&&\hskip -2cm
 +\frac3{4}H^2aa'\ln(aa')\partial^2\theta(\Delta\eta\!-\!r)
 + H^4a^2{a'}^2\bigg[\frac{3}{4\!-\!y}\bigg]\theta(\Delta\eta\!-\!r)
    \Bigg\}
 \,.
\qquad
\label{F21+G21}
\end{eqnarray}
The equation for the gravitational slip~(\ref{one loop eom delta ij}) then becomes,
\begin{eqnarray}
\nabla^2(f_3^{(1)}\!-\!f_1^{(1)})=\frac{\kappa^2M}{a^2}\nabla^2S_0^{(1)}(x)
+\frac{\kappa^2M}{a^2}\Big(\!-\partial_{0}^2\!+\!\frac13\nabla^2\Big) S_2^{(1)}(x)
\,.\quad
\label{one loop eom: gravitational slip}
\end{eqnarray}
The solutions for $f_1^{(1)}$ and $f_3^{(1)}$ are obtained by combining~(\ref{one loop  eom 00}) and~(\ref{one loop  eom 0i}),
\begin{eqnarray}
f_1^{(1)}(x) &=& -\frac{\kappa^2M}{2a^2} S_0^{(1)}(x)
+\frac{\kappa^2M }{a^2}\Big[\!-\!\frac23\!+\!\nabla^{-2}(\partial_{0}^2\!-\!aH\partial_0)\Big]S_2^{(1)}(x)\equiv-2\phi^{(1)}(x)
\;, \quad
\label{one loop eom: f1}
\\
f_3^{(1)}(x) &=& \frac{\kappa^2M}{2a^2} S_0^{(1)}(x)
+\frac{\kappa^2M }{a^2}\Big[\!-\!\frac13\!-\!\nabla^{-2}aH\partial_0\Big]S_2^{(1)}(x)\equiv-2\psi^{(1)}(x)
\;,\quad
\label{one loop eom: f3}
\end{eqnarray}
where 
\be
\nabla^{-2}f(\eta,\vec x) = -\frac{1}{4\pi} \int d^3 x' \frac{f(\eta,\vec x')}{\|\vec x-\vec x'\|} \;.
\ee
By inserting~(\ref{one loop eom: f1}) and~(\ref{one loop eom: f3})
into~(\ref{one loop eom 00}--\ref{one loop eom partial ij}) one sees that all of
the equations~(\ref{one loop eom 00}--\ref{one loop eom partial ij}) are satisfied,
representing a nontrivial check of our basic equations~(\ref{one loop eom: f1}--\ref{one loop eom: f3}).
We also see that the spatial integral of the spin two source
is required for determination of the one-loop contributions to
both gravitational potentials $\phi^{(1)}=-f_1^{(1)}/2$ and $\psi^{(1)}=-f_3^{(1)}/2$.

 The actual calculation of the quantum (one-loop) corrected gravitational potentials $\phi^{(1)}$ and $\psi^{(1)}$
is rather technical and we relegate it to the Appendix. Since the complete results are rather complex, in the last step of the calculation in the Appendix we take the late time limiting
form. The final result for $f_1^{(1)}$ and  $f_3^{(1)}$ is given in Eqs.~(\ref{f1:final}--\ref{f3:final}). From these, it is easy to extract the one-loop 
corrected potentials $\phi^{(1)} = -f_1^{(1)}/2$, $\psi^{(1)}  = -f_3^{(1)}/2$. When 
$\kappa$ is replaced with $G$ ({\it via} the relation, $\kappa=\sqrt{16\pi G}$) and 
 the units $c, \hbar$ are appropriately re-inserted to elucidate the quantum gravitational nature
of our calculation, we obtain from~(\ref{f1:final}--\ref{f3:final}), 
\begin{eqnarray}
\phi^{(1)} (x) &\!\!=\!\!& -\frac{GM}{ar}\Biggl\{
  \frac{\hbar}{20\pi c^3}\frac{G}{(ar)^2}
+\frac{\hbar}{\pi c^5} GH^2\biggl[
   -\frac{1}{3}\ln(a) -\frac{3}{10}\ln\Bigl(\frac{Hr}{c}\Bigr)
                       +{\cal O}\Big(\frac{1}{a^3}\Big)\biggr]
            \Biggr\}
\label{phi:final}\\
\psi^{(1)}(x) &\!\!=\!\!& -\frac{GM}{ar}\Biggl\{
 -\frac{\hbar}{60\pi c^3}\frac{G}{(ar)^2}
+\frac{\hbar}{\pi c^5} GH^2\biggl[
  -\frac{1}{3}\ln(a) -\frac{3}{10}\ln\Bigl(\frac{Hr}{c}\Bigr) +\frac{2}{3}\frac{Har}{c}
                           \!+\!{\cal O}\Big(\frac{1}{a^3}\Big)\biggr]
            \Biggr\}
\,.\qquad
\label{psi:final}
\end{eqnarray}
It follows that MMC scalars in de Sitter background generate the conformal scalar contributions 
plus another positive contributions to the gravitational potentials. 
The first terms in Eqs.~(\ref{phi:final}--\ref{psi:final}) represent the one-loop contributions from a conformal scalar field,
and in the limit when $H\rightarrow 0, a\rightarrow 1$ they reduce to the Minkowski space results of Refs. \cite{PWforceflat, MP},
representing a nontrivial check of our principal results~(\ref{phi:final}--\ref{psi:final}).


\section{Discussion}\label{discuss}

We have included one loop effects from MMC scalars to derive quantum loop 
corrections to the potentials  associated with a static point mass.   
Each of the full potentials, $\phi$ and $\psi$, can be presented as its classical value 
times a series of quantum corrections we have obtained in (\ref{phi:final}--\ref{psi:final}), 
which are at late times (when $a\gg 1$),
\bea
\phi_{\rm dS}(x) &\!\!=\!\!& -\frac{GM}{ar}\Biggl\{\!1
+\frac{\hbar}{20\pi c^3}\frac{G}{(ar)^2}
+\frac{\hbar G H^2}{\pi c^5} \biggl[
  -\frac{1}{30}\ln(a) -\frac{3}{10}\ln\Bigl(\frac{Har}{c}\Bigr)
                       \biggr] +{\cal O}\Big(G^2,\frac{1}{a^3}\Big) \!
            \Biggr\}
\label{phi:full-dS}\\
\psi_{\rm dS}(x) &\!\!=\!\!& -\frac{GM}{ar}\Biggl\{\!1
 -\frac{\hbar}{60\pi c^3}\frac{G}{(ar)^2}
+\frac{\hbar G H^2}{\pi c^5} \biggl[
 \! -\frac{1}{30}\ln(a) -\frac{3}{10}\ln\Bigl(\frac{Har}{c}\Bigr) +\frac{2}{3}\frac{Har}{c}
                        \biggr]   +{\cal O}\Big(G^2,\frac{1}{a^3}\Big)\! 
            \Biggr\}
.\nonumber\\
\label{psi:full-dS}
\eea
These can be compared with the corresponding flat space results which have been previously computed 
in \cite{PWforceflat, MP},
\bea
\phi_{\rm flat}(x) &\!\!=\!\!& -\frac{GM}{r}\Biggl\{1
+\frac{\hbar}{20\pi c^3}\frac{G}{r^2}
                       +{\cal O}(G^2)            
                       \Biggr\}
\label{phi:full-flat}\\
\psi_{\rm flat}(x) &\!\!=\!\!& -\frac{GM}{r}\Biggl\{1
 -\frac{\hbar}{60\pi c^3}\frac{G}{r^2}
                           +{\cal O}(G^2)
            \Biggr\}
\,.
\label{psi:full-flat}
\eea
From this comparison, one can see that the first quantum correction terms in~(\ref{phi:full-dS}--\ref{psi:full-dS}) represent the de Sitter version of the flat space correction and the terms proportional to $GH^2$ are the intrinsic de Sitter corrections. 

Note that every factor of the co-moving distance $r$ which appears in expressions
(\ref{phi:full-dS}-\ref{psi:full-dS}) is multiplied by a scale factor $a(\eta)$ so
that their product gives the physical distance from source to observation point. 
The remaining factor of $\ln(a)$ multiplies a term of the same form as the classical
potential. Because these secular terms contribute equally to both potentials, 
they can be reinterpreted as a time dependent renormalization of the mass term,
\begin{equation}
   M\rightarrow M\bigg[1-\frac{\hbar}{c^5}\frac{GH^2}{30\pi}\ln(a)\bigg] 
\,,
\end{equation}
or equivalently a time dependent renormalization of the Newton's constant,
\begin{equation}
   G\rightarrow G\bigg[1-\frac{\hbar}{c^5}\frac{GH^2}{30\pi}\ln(a)\bigg]
\,.
\end{equation}
Even though the secular terms are suppressed by the loop counting parameter, 
$\hbar GH^2/c^5$, whose value is less than $10^{-10}$ for primordial inflation, 
they are growing in time, and can eventually become large. Indeed, when the 
number of e-foldings, $\ln(a)=Ht$ becomes of the order $c^5/(\hbar GH^2)$, the 
correction becomes large, signifying a breakdown of perturbation 
theory in the sense that, when $[\hbar GH^2/c^5]\ln(a)\sim 1$, all orders contribute 
significantly. To understand what happens at very late times, one would have to sum
these higher loop contributions, which is a major unsolved problem~\cite{mw,
Kitamoto:2010et,Kitamoto:2011yx}.

The secular screening effect we have just described is fascinating. It might
represent the seed of an explanation for why the Newton constant seems so much
smaller than any other length scale of fundamental theory. However, there is
no avoiding the sense of strangeness. If we adopt the perspective of an 
observer at fixed co-moving position, whose physical distance to the source 
increases exponentially in co-moving time, then quantum scalar fluctuations are
erasing the gravitational imprint of a point source faster than its classical
redshift, in precisely the region where the source has almost no effect. From 
the perspective of an observer at fixed physical distance (in static coordinates) 
one wonders why anything is changing at all.

The sense of strangeness is even stronger when we compare with the recent 
result \cite{WW} for one loop corrections to the gravitational potentials 
from virtual photons. Unlike our case of MMC scalars, conformal invariance
means that photons behave the same, in de Sitter conformal coordinates, as
they do in flat space. Yet they also induce secular screening \cite{WW}. 
From this we can infer that the screening effect originates not so much from 
the way quantum fluctuations are affected by inflation but rather from the 
different way that gravity responds to sources on de Sitter space as opposed 
to flat space.  

Finally, it is interesting to speculate that quantum corrections to gravity
from the epoch of primordial inflation might modify late time gravity in
observable ways, for example, as regards explaining the current phase of
cosmic acceleration. These type of questions have been investigated in the 
context of Einstein's gravity endowed with a non-minimally coupled, light
scalar~\cite{Glavan:2014uga,Glavan:2015}, we well as in the context of some 
simple non-local extensions of gravity~\cite{DW,DP}.

\vskip 1cm

\centerline{\bf Acknowledgements}

We are grateful for conversations on this subject with M. Fr\"ob and E. Verdaguer. 
This work was partially supported by KASI, by the D-ITP consortium, a program of the 
NWO that is funded by the Dutch Ministry of Education, Culture and Science (OCW), 
by NSF grants PHY-1205591 and PHY-1506513, and by the Institute for Fundamental 
Theory at the University of Florida.


\section*{Appendix}

In this Appendix we give some details on how to perform the source integrals ${S}_0^{(1)}$ and  ${S}_2^{(1)}$
given in Eqs.~(\ref{S01}--\ref{S21}) and~(\ref{F01}--\ref{F21+G21}),  which through 
Eqs.~(\ref{one loop eom: f1}--\ref{one loop eom: f3})
allow us to calculate the one-loop corrected scalar gravitational potentials  
$\phi^{(1)}=-f^{(1)}_1/2$ and $\psi^{(1)}=-f_3^{(1)}/2$. 

To begin with, it is convenient to break the source integrals, $S_0^{(1)}$ and $S_2^{(1)}$ 
in Eqs.~(\ref{S01}--\ref{S21}) into the following simpler integrals,
\begin{eqnarray}
{\cal I}_{1a}&\equiv& 
\frac{1}{16}\partial^4\int_{\eta_0=-1/H}^{\eta-r}\frac{d\eta'}{a'}\left[\ln\bigg(\frac{-y}{4aa'}\bigg)\!-\!1\right]
 = \frac{1}{4ar^3}
 \label{I1a}\\
 {\cal I}_{1b}&\equiv& 
 -\frac{H^2a}{4}\partial^2\int_{\eta_0=-1/H}^{\eta-r}\!\!d\eta'\left[\ln\bigg(\frac{-y}{4aa'}\bigg)\!-\!1\right]
 =\frac{H^2a}{r}\Bigg\{ 
 \ln(Hr)\!+\!\frac12\ln\bigg(\frac{1\!-\!Hr\!-\!\frac1a}{1\!+\!Hr\!-\!\frac1a}\bigg)
 \Bigg\}
 \label{I1b}\\
 {\cal I}_{2}&\equiv&
 -\frac{H^2a}{4}\bigg[\ln(a)\partial^2\!\int_{\eta_0=-1/H}^{\eta-r}\!\!d\eta'\!+\!\partial^2\!\int_{\eta_0=-1/H}^{\eta-r}d\eta'\ln(a')\bigg]
 =\frac{H^2a}{2r} \ln\bigg(\frac{a}{Hr\!+\!\frac1a}\bigg)
 \label{I2}\\
 {\cal I}_{3a}&\equiv& H^4a^2\int_{\eta_0=-1/H}^{\eta-r}\!\!d\eta'a' \left[-\frac{1}{4-y}\right]
 = - \frac{H^2a}{2r} \Bigg\{ 
 \ln(1\!+\! aHr)\!+\!\ln\bigg(\frac{1 \!-\! Hr \!+\! \frac1a}{1 \!+\!Hr\!+\! \frac1a }\bigg)
 \Bigg\}
 \label{I3a}\\
 {\cal I}_{3b}&\equiv& 3H^4a^2\int_{\eta_0=-1/H}^{\eta-r}\!\!d\eta'a'\left[1\!+\!\frac14(2\!-\!y)\ln\bigg(\frac{-y}{4\!-\!y}\bigg)\right]
 =\frac{3}{4}H^3a^3\Bigg\{
 4Hr\ln\bigg(1\!+\!\frac1{Har}\bigg)
\nonumber\\
&& -\bigg[\!(1\!+\!Hr)^2\!-\!\frac{1}{a^2} \!\bigg]\ln\bigg(\frac{1 \!+\!Hr \!+\! \frac1a }{1\!+\!Hr \!-\! \frac1a }\bigg)
\!-\!\bigg[\!(1\!-\!Hr)^2\!-\!\frac{1}{a^2} \!\bigg]\ln\bigg(\frac{1 \!-\!Hr\!+\! \frac1a }{1 \!-\!Hr \!-\! \frac1a }\bigg)\Bigg\}
\,.
 \label{I13b}
\end{eqnarray}
When written in terms of these integrals, the spin zero and spin two sources~(\ref{S01}--\ref{S21}) and~(\ref{F01}--\ref{F21+G21})
are simply,
\begin{eqnarray}
S^{(1)}_0(t,r) &=&\!\! -\frac{\kappa^2}{64\times 3\pi^3}\Big[
       \frac13{\cal I}_{1a}\!+\!\frac13 ({\cal I}_{1b}\!+\! {\cal I}_{2}\!+\! {\cal I}_{3a})\!+\! \frac13{\cal I}_{3b} \Big]
\label{S01:b}
\\
S^{(1)}_2(t,r) &=&\!\! -\frac{\kappa^2}{64\times 3\pi^3}\bigg[
       \frac15{\cal I}_{1a}\!-\! ({\cal I}_{1b}\!+\! {\cal I}_{2}\!+\! {\cal I}_{3a}) \bigg]
\,,
\quad
\label{S21:b}
\end{eqnarray}
where
\begin{equation}
{\cal I}_{1b}\!+\! {\cal I}_{2}\!+\! {\cal I}_{3a} =
-\frac{H^2a}{r}\Biggl\{ \ln\Bigl[1\!+\!\frac1{Har}\Bigr]\!
+\!\frac12 \ln\biggl[\frac{(1 \!+\!Hr \!-\! \frac1a )(1 \!-\!Hr\!+\! \frac1a )}{(1 \!+\!Hr \!+\! \frac1a )(1 \!-\!Hr \!-\! \frac1a )}\biggr] \Biggr\}
\,.
\label{I1b+I2+I3a}
\end{equation}

Now, it is convenient to re-express the scalar gravitational fields $f_1^{(1)}$ and $f_3^{(1)}$ in
Eqs.~(\ref{one loop eom: f1}--\ref{one loop eom: f3}) as,
\begin{eqnarray}
f_1^{(1)}(x) &=& \kappa^2M\bigg[-\frac{1}{2a^2} S_0^{(1)}(x)-\frac{2}{3a^2}S_2^{(1)}(x)+\partial_{t}^2\nabla^{-2}S_2^{(1)}(x)
            \bigg]
\label{f1:b}\\
f_3^{(1)}(x) &=& \kappa^2M\bigg[\frac{1}{2a^2} S_0^{(1)}(x)-\frac{1}{3a^2}S_2^{(1)}(x)-H\partial_t\nabla^{-2}S_2^{(1)}(x)
            \bigg]
\,,
\label{f3:b}
\end{eqnarray}
where
\begin{eqnarray}
 S_0^{(1)} &=&  -\frac{\kappa^2}{64\times 3\pi^3}\Biggl\{
       \frac1{12}\frac{a^2}{(ar)^3}+a^2H^3\Biggl[
       -\frac{1}{3Har}\ln\Big(1\!+\!\frac1{Har}\Big)+Har\ln\Big(1\!+\!\frac1{Har}\Big)
 \nonumber\\
    && \hskip 0.5cm
  +\frac{1}{6Har}
              \ln\bigg(\frac{(1 \!+\!Hr\!+\! \frac1a )(1\!-\!Hr\!-\! \frac1a )}{(1\!+\!Hr\!-\! \frac1a )(1\!-\!Hr\!+\! \frac1a )}\bigg)
    \!- \frac{a}4\bigg((1\!+\!Hr)^2\!-\!\frac{1}{a^2}\bigg)\ln\bigg(\frac{1\!+\!Hr\!+\! \frac1a }{1\!+\!Hr\!-\! \frac1a }\bigg)
     \nonumber\\
     && \hskip 0.5cm
\!-\frac{a}4\bigg((1\!-\!Hr)^2\!-\!\frac{1}{a^2}\bigg)\ln\bigg(\frac{1\!-\!Hr\!+\! \frac1a }{1\!-\!Hr\!-\! \frac1a }\bigg)
\Biggr]
\Biggr\}
\,\quad
\label{s01:c}
\\
 S_2^{(1)} &=&  -\frac{\kappa^2}{64\times 3\pi^3}\Biggl\{
       \frac1{20}\frac{a^2}{(ar)^3}
\nonumber \\
   &&\hskip 0.5cm
    +a^2H^3\Biggl[\frac{1}{Har}\ln\Big(1\!+\!\frac1{Har}\Big)
    \!+\!\frac{1}{2Har}
              \ln\bigg(\frac{(1\!+\!Hr\!-\! \frac1a )(1\!-\!Hr\!+\! \frac1a )}{(1\!+\!Hr\!+\! \frac1a )(1\!-\!Hr\!-\! \frac1a )}\bigg)\Biggr]
\Biggr\}
\,.
\label{s21:c}
\end{eqnarray}
To fully reconstruct the gravitational potentials~(\ref{f1:b}--\ref{f3:b}) 
we also need to evaluate $\nabla^{-2}S_{2}^{(1)}$, which can be broken into three parts,
\begin{eqnarray}
   \nabla^{-2}S_{2}^{(1)} &=& -\frac{\kappa^2H^3}{64\times 3\pi^3}\Biggl\{
       \frac1{20aH^3}\nabla^{-2}\bigg(\frac{1}{r^3}\bigg)
\nonumber
\\
   &&\hskip 0.5cm
    +\nabla^{-2}\bigg[\frac{a}{Hr}\ln\Big(1\!+\!\frac1{Har}\Big)\bigg]
    \!+\!\nabla^{-2}\bigg[\frac{a}{2Hr}
               \ln\bigg(\frac{(1\!+\!Hr\!-\! \frac1a )(1\!-\!Hr\!+\! \frac1a )}{(1\!+\!Hr\!+\! \frac1a )(1\!-\!Hr\!-\! \frac1a )}\bigg)
              \bigg]\Biggr\}
\,.\quad
\label{s21: nable}
\end{eqnarray}
There are three pieces on which the inverse Laplace operator act. One can check that the first evaluates to,
\begin{equation}
  {\cal J}_1\equiv\frac1{20aH^3}\nabla^{-2}\bigg(\frac{1}{r^3}\bigg) = -\frac1{20H^3}\frac{\ln(Hr)}{ar}
\,,
\label{Laplace 1}
\end{equation}
where, for convenience, we fixed the integration constant to 
be $-\ln(H)/r$.~\footnote{Other choices of integration constants are possible. The plausible one is,
$-\ln(Ha)/r$, since in this case the final answer depends on the physical distance $ar$ only. 
However, an inspection of the constraint equation~(\ref{one loop eom partial ij}) shows that this choice is illegitimate.
Indeed, since ${\cal J}_1$ contributes equally to both $f_1$ and $f_3$, but it does not contribute to the right hand side,
${\cal J}_1$ must satisfy, $(\partial_0^2+3aH\partial_0+3a^2H^2){\cal J}_1=a^2(\partial_t^2+4H\partial_t+3H^2){\cal J}_1=0$, 
from which we conclude
 that the choice of the integration constant in~(\ref{Laplace 1}) is the correct one.  
}
Strictly speaking, when $\nabla^2$ acts on $-\ln(H)/r$ it generates a delta function, $\propto \delta^3(\vec x\,)$,
but that term can be subsumed in a (finite) renormalization of the Newton constant.
To evaluate the other two terms in~(\ref{s21:c}) the following integral representation can be used,
\begin{equation}
  \nabla^{-2} g(r) = \int_0^r dr'r' \Big(1-\frac{r'}{r}\Big)g(r')
  \,,
\label{inverse Laplace:rule}
\end{equation}
where $[r^2g(r)]_{r\rightarrow 0}$ must vanish.
Applying this to the other two terms in~(\ref{s21:c}) yields,
\bea
{\cal J}_2 &\equiv& \nabla^{-2}\bigg[\frac{a}{Hr}\ln\Big(1\!+\!\frac1{Har}\Big)\bigg]
\nn\\
&=& \frac{a}{2H^3r}\Big(Hr+\frac1a\Big)^2\bigg[\ln\Big(Hr\!+\!\frac{1}{a}\Big)\!-\!\frac32\bigg]
      -\frac{ar}{2H}\bigg[\ln(Hr)\!-\!\frac32\bigg]
      +\frac{1}{H^2}\bigg[\ln(a)\!+\!1\bigg]
      \nn\\
&& \hskip 0.5cm      
      +\frac{1}{2raH^3}\bigg[\ln(a)\!+\!\frac32\bigg]
\label{s21:d1}
\\
{\cal J}_3 &\equiv&  \nabla^{-2}\bigg[\frac{a}{2Hr}
               \ln\bigg(\frac{(1\!+\!Hr\!-\! \frac1a )(1\!-\!Hr\!+\! \frac1a )}{(1\!+\!Hr\!+\! \frac1a )(1\!-\!Hr\!-\! \frac1a )}\bigg)
              \bigg]
\nn\\
&=& \frac{a}{4H^3r}\bigg\{\Big(1\!+\!Hr\!-\!\frac1a\Big)^2\bigg[\ln\Big(1\!+\!Hr\!-\!\frac1a\Big)\!-\!\frac32\bigg]
+\Big(1\!-\!Hr\!+\!\frac1a\Big)^2\bigg[\ln\Big(1\!-\!Hr\!+\!\frac1a\Big)\!-\!\frac32\bigg]
\nn\\
&& \hskip 1cm
-\Big(1\!+\!Hr\!+\!\frac1a\Big)^2\bigg[\ln\Big(1\!+\!Hr\!+\!\frac1a\Big)\!-\!\frac32\bigg]
-\Big(1\!-\!Hr\!-\!\frac1a\Big)^2\bigg[\ln\Big(1\!-\!Hr\!-\!\frac1a\Big)\!-\!\frac32\bigg]
\bigg\}
\nn\\
&& \hskip 1cm
+\frac{a}{H^2}\Big(1\!+\!\frac1a\Big)\bigg[\ln\Big(1\!+\!\frac1a\Big)\!-\!1\bigg]
      -\frac{a}{H^2}\Big(1\!-\!\frac1a\Big)\bigg[\ln\Big(1\!-\!\frac1a\Big)\!-\!1\bigg]
\;.
\label{s21:d2}
\eea
In terms of these integrals, $\nabla^{-2}S_2^{(1)}$ is simply,
\begin{equation}
 \nabla^{-2}S_2^{(1)} = -\frac{\kappa^2H^3}{64\times 3\pi^3}\Big(
       {\cal J}_1 + {\cal J}_2 + {\cal J}_3\Big)
\,.\quad
\label{s21:e}
\end{equation}

According to the expressions for the scalar potentials $f_1^{(1)}$ and $f_3^{(1)}$ in (\ref{f1:b}--\ref{f3:b}) we need the following combinations of the sources $S_0^{(1)}$ and $S_2^{(1)}$, 
%
\bea
\lefteqn{-\frac{S_0^{(1)}}{2a^2}-\frac{2S_2^{(1)}}{3a^2} =  \frac{\kappa^2}{128\times 3\pi^3}\Biggl\{
       \frac3{20}\frac{1}{(ar)^3}+H^3\bigg[\frac{1}{Har}\ln\Big(1\!+\!\frac1{Har}\Big)+Har\ln\bigg(1\!+\!\frac1{Har}\bigg)}
 \nn\\
    && \hskip 0.5cm +\frac{1}{2Har}
              \ln\bigg(\frac{(1\!+\!Hr\!-\! \frac1a)(1\!-\!Hr\!+\! \frac1a)}{(1\!+\!Hr\!+\! \frac1a)(1\!-\!Hr\!-\! \frac1a)}\bigg)
    \!- \frac{a}4\bigg((1\!+\!Hr)^2\!-\!\frac{1}{a^2}\bigg)\ln\bigg(\frac{1\!+\!Hr\!+\! \frac1a}{1\!+\!Hr\!-\! \frac1a}\bigg)
     \nn\\
     && \hskip 0.5cm
-\frac{a}4\bigg((1\!-\!Hr)^2\!-\!\frac{1}{a^2}\bigg)\ln\bigg(\frac{1\!-\!Hr\!+\! \frac1a}{1\!-\!Hr\!-\! \frac1a}\bigg)\bigg]\Biggr\}
\,,\quad \quad
\label{s01+s21:a}
\\
\lefteqn{\frac{S_0^{(1)}}{2a^2}-\frac{S_2^{(1)}}{3a^2}  = \frac{\kappa^2}{128\times 3\pi^3}\Biggl\{
       \!-\frac1{20}\frac{1}{(ar)^3}+H^3\bigg[\frac{4}{3Har}\ln\Big(1\!+\!\frac1{Har}\Big)-Har\ln\bigg(1\!+\!\frac1{Har}\bigg)}
 \nn\\
    && \hskip 0.5cm +\frac{1}{2Har}
              \ln\bigg(\frac{(1\!+\!Hr\!-\! \frac1a)(1\!-\!Hr\!+\! \frac1a)}{(1\!+\!Hr\!+\! \frac1a)(1\!-\!Hr\!-\! \frac1a)}\bigg)
    \!+ \frac{a}4\bigg((1\!+\!Hr)^2\!-\!\frac{1}{a^2}\bigg)\ln\bigg(\frac{1\!+\!Hr\!+\! \frac1a}{1\!+\!Hr\!-\! \frac1a}\bigg)
     \nn\\
     && \hskip 0.5cm
+\frac{a}4\bigg((1\!-\!Hr)^2\!-\!\frac{1}{a^2}\bigg)\ln\bigg(\frac{1\!-\!Hr\!+\! \frac1a}{1\!-\!Hr\!-\! \frac1a}\bigg)\bigg]\Biggr\}
\,.\quad \quad
\label{s01+s21:b}
\eea

\subsection{Late time limit of the gravitational potentials}

We now have all the ingredients to calculate $f_1^{(1)}$ and $f_3^{(1)}$ given in (\ref{f1:b}--\ref{f3:b}). The answer is rather long, and since we are primarily interested in 
the late time behavior, we now present the late time limit, $a\rightarrow \infty$, of various relevant contributions.
First for the non-spatial integral terms in (\ref{f1:b}--\ref{f3:b}) we have
\begin{eqnarray}
 -\frac{S_0^{(1)}}{2a^2}\!-\!\frac{2S_2^{(1)}}{3a^2}
 &\!\!\stackrel{a\rightarrow \infty}{\longrightarrow}\!\!&
  \frac{\kappa^2H^3}{64\times 3\pi^3}
 \bigg[\frac{3}{40(Har)^3}\!-\!\frac{1}{4Har}\!+\!\frac{2}{3(Har)^2}\!-\!\frac{3}{8(Har)^3}\!+\!\frac{4}{3(1\!-\!H^2r^2)a^2}
                       \!+\!{\cal O}\Big(\frac{1}{a^4}\Big)\bigg]
\nonumber\\
\label{S01:late time}\\
   \frac{S_0^{(1)}}{2a^2}\!-\!\frac{S_2^{(1)}}{3a^2} 
  &\!\!\stackrel{a\rightarrow \infty}{\longrightarrow}\!\!&
  \frac{\kappa^2H^3}{64\times 3\pi^3}
 \bigg[\frac{-1}{40(Har)^3}\!+\!\frac{1}{4Har}\!+\!\frac{1}{3(Har)^2}\!-\!\frac{1}{8(Har)^3}\!+\!\frac{2}{3(1\!-\!H^2r^2)a^2}
                           \!+\!{\cal O}\Big(\frac{1}{a^4}\Big)\bigg]
 \,,\nonumber\\
\label{S21:late time}
\end{eqnarray}
For the spatial integral terms (\ref{s21:d1}--\ref{s21:d2}) which eventually enter (\ref{f1:b}--\ref{f3:b}), we obtain 
\begin{eqnarray}
{\cal J}_2&\stackrel{a\rightarrow \infty}{\longrightarrow}&\frac{1}{H^2}
 \bigg\{\ln(Har)\!+\!\frac{1}{2Har}\bigg[\ln(Har)\!+\!\frac32\bigg]\!+\!\frac{1}{6(Har)^2}\!-\!\frac{1}{24(Har)^3}
             \!+\!{\cal O}\Big(\frac{1}{a^4}\Big)\bigg\}
\label{J2:late time}\\
{\cal J}_3 &\stackrel{a\rightarrow \infty}{\longrightarrow}&\frac{1}{H^2}
 \bigg\{\frac{1\!+\!Hr}{Hr}\ln(1\!+\!Hr)\!+\!\frac{1\!-\!Hr}{Hr}\ln(1\!-\!Hr)\!+\!2\!+\!\frac{1}{3(1\!-\!H^2r^2)a^2}
             \!+\!{\cal O}\Big(\frac{1}{a^4}\Big)\bigg\}
 \,.
\label{J3:late time}
\end{eqnarray}
Then the action of the inverse Laplacian operator on the spin two source becomes in the large $a$ limit,
\begin{eqnarray}
 \lim_{a\rightarrow \infty} \nabla^{-2}S_2^{(1)} 
 &=& - \lim_{a\rightarrow \infty} \frac{\kappa^2H^3}{64\times 3\pi^3}\Big(
       {\cal J}_1 + {\cal J}_2 + {\cal J}_3\Big)
       \nn\\
 &=&
  \frac{\kappa^2 H}{64\times 3\pi^3}
  \bigg\{\frac{\ln(Hr)}{20Har}
   \!-\ln(Har)\!-\!\frac{1}{2Har}\bigg[\ln(Har)\!+\!\frac32\bigg]\!-\!\frac{1}{6(Har)^2}\!+\!\frac{1}{24(Har)^3}
  \nonumber\\
  &&\!-\frac{1\!+\!Hr}{Hr}\ln(1\!+\!Hr)\!-\!\frac{1\!-\!Hr}{Hr}\ln(1\!-\!Hr)\!-\!2\!-\!\frac{1}{3(1\!-\!H^2r^2)a^2}
             \!+\!{\cal O}\Big(\frac{1}{a^4}\Big)\bigg\}
 \,,
\label{inv laplace S21:late time}
\end{eqnarray}
What enters $f_1^{(1)}$ and $f_3^{(1)}$ are the second and first derivative of this expression, respectively, {\it i.e.}
\begin{eqnarray}
 \partial_t^2\nabla^{-2}S_2^{(1)} &\stackrel{a\rightarrow \infty}{\longrightarrow}&\frac{\kappa^2 H^3}{64\times 3\pi^3}
  \bigg\{\frac{\ln(Hr)}{20Har}
   \!-\!\frac{1}{2Har}\bigg[\ln(Har)\!-\!\frac12\bigg]\!-\!\frac{2}{3(Har)^2}\!+\!\frac{3}{8(Har)^3}
  \nonumber\\
  &&\!-\frac{4}{3(1\!-\!H^2r^2)a^2}
             \!+\!{\cal O}\Big(\frac{1}{a^4}\Big)\bigg\}
 \,,
\label{dt2 inv laplace S21:late time}\\
 -H\partial_t\nabla^{-2}S_2^{(1)} &\stackrel{a\rightarrow \infty}{\longrightarrow}&\frac{\kappa^2 H^3}{64\times 3\pi^3}
  \bigg\{\frac{\ln(Hr)}{20Har}
   \!+1\!-\!\frac{1}{2Har}\bigg[\ln(Har)\!+\!\frac12\bigg]\!-\!\frac{1}{3(Har)^2}\!+\!\frac{1}{8(Har)^3}
  \nonumber\\
  &&\!-\frac{2}{3(1\!-\!H^2r^2)a^2}
             \!+\!{\cal O}\Big(\frac{1}{a^4}\Big)\bigg\}
 \,.
\label{dt inv laplace S21:late time}
\end{eqnarray}
Interestingly, all the negative powers of $a$ without the logarithm factors in (\ref{dt2 inv laplace S21:late time}--\ref{dt inv laplace S21:late time}) cancel the corresponding terms in (\ref{S01:late time}--\ref{S21:late time}).  
What finally remains in the scalar perturbations $f_1^{(1)}$ and $f_3^{(1)}$ at late times are,
\begin{eqnarray}
f_1^{(1)}(x) &=& \kappa^2M\frac{\kappa^2H^3}{64\times 3\pi^3}\bigg[
  \frac{3}{40(Har)^3}\!+\!\frac{\ln(Hr)}{20Har}
   \!-\!\frac{\ln(Har)}{2Har}
                       \!+\!{\cal O}\Big(\frac{1}{a^4}\Big)
            \bigg]
\label{f1:final}\\
f_3^{(1)}(x) &=& \kappa^2M\frac{\kappa^2H^3}{64\times 3\pi^3}\bigg[
  \!-\!\frac{1}{40(Har)^3}\!+\!\frac{\ln(Hr)}{20Har}
   \!+\!1\!-\!\frac{\ln(Har)}{2Har}
                           \!+\!{\cal O}\Big(\frac{1}{a^4}\Big)
            \bigg]
\,.
\label{f3:final}
\end{eqnarray}
These are our main results, which are used in the main text~(\ref{phi:final}--\ref{psi:final}) 
to obtain the late time one-loop corrected gravitational potentials.

\end{document}